\documentclass[floatfix,onecolumn,pre]{revtex4}
\usepackage[pdftex]{graphicx}
\usepackage{amssymb}
\usepackage{amsmath}
\usepackage{latexsym}
\usepackage{bm}
\usepackage{times,subfigure}
\newcommand{\young}{\ensuremath{\theta_{\mathrm{Y}}}}

\newcommand{\youni}{\ensuremath{\theta_{\mathrm{I}}}}

\begin{document}
\title{Anisotropic imbibition on surfaces patterned with polygonal posts}
\author{MATTHEW L. BLOW{\footnote{Now at: Centro de F\'{i}sica Te\'{o}rica e Computacional (CFTC), University of Lisbon, Instituto de Investiga\c{c}\~{a}o Interdisciplinar, Av. Prof. Gama Pinto, 2, Lisboa, 1649-003, Portugal}} {\footnote{{\bf email:} matthewlblow@gmail.com}}}
\author{JULIA M. YEOMANS}
\affiliation{The Rudolf Peierls Centre for Theoretical Physics, Oxford University, 1 Keble Road, Oxford OX1 3NP, England}

\begin{abstract}
 We present and interpret lattice Boltzmann simulations of thick films spreading on 
surfaces patterned with polygonal posts. We show that the mechanism of pinning and 
depinning differs with the direction of advance, and demonstrate that this leads to 
anisotropic spreading within a certain range of material contact angles.
\end{abstract}

\maketitle

\section{Introduction}
\label{intro}
A drop placed on a partially wetting substrate will make a finite angle with the 
surface, given by Young's equation~\cite{Young},
\begin{equation}
\cos\young=\frac{\gamma_{\mathrm{SV}}-\gamma_{\mathrm{SL}}}{\gamma}\;,  
\label{eqn:young}
\end{equation}
where $\gamma_{\mathrm{SV}}$, $\gamma_{\mathrm{SL}}$ and ${\gamma}$ are the 
solid--vapour, solid--liquid and liquid--vapour surface tensions. Young's equation 
assumes that the surface is smooth, and that the contact line is able to move freely 
to allow the drop to globally minimise its free energy.

Due to recent advances in microlithography it is now possible to pattern surfaces with 
regular arrays of micron-scale posts, leading to deviations from Young's equation on a 
macroscopic level. On a superhydrophilic, or superwetting surface, the fluid can be 
drawn into the spaces between the posts, such that the drop forms a film with 
thickness equal to the height of the posts \cite{Bico,Ishino,Ishino2}, a phenomenon 
that is termed imbibition.

Imbibition is thermodynamically feasible if the thick film has a lower free energy 
than the dry surface. The free energy change per unit width, $\delta \mathcal{F}$, 
when the film advances a distance $\delta x$ can be estimated by averaging over the 
surface features
\begin{equation}
\delta\mathcal{F}=\left[\left(\gamma_{SL}-\gamma_{SV}\right)\left(r-\phi\right)+\gamma
\left(1-\phi\right)\right]\delta x\;,  \label{eqn:freeEnergyChange}
\end{equation}
where $r$ is the ratio of the surface area to its vertical projection,  $\phi$ is the 
fraction of the surface covered by posts, and we assume posts of constant 
cross-section. Eliminating the surface tensions in Eqn.~(\ref{eqn:freeEnergyChange}) 
using Eqn.~(\ref{eqn:young}), the condition $\delta\mathcal{F}<0$ becomes~\cite{Bico}
\begin{equation}
\cos\young>\cos\youni=\frac{1-\phi}{r-\phi}\;.   \label{eqn:bico}
\end{equation}
This inequality relies on the same assumption as Young's equation (\ref{eqn:young}):
namely that the contact line can move freely over the substrate, sampling the average 
properties of the roughness. This assumption holds well on some surfaces, for example, 
in the longitudinal direction on a grooved surface, but not in other cases, for 
example perpendicular to such grooves, where free energy barriers due to contact line 
pinning can halt the motion of the interface~\cite{ChenHe}.

\begin{figure}
\centering
\subfigure[]{\label{fig:gibbs}\includegraphics[width=50mm]{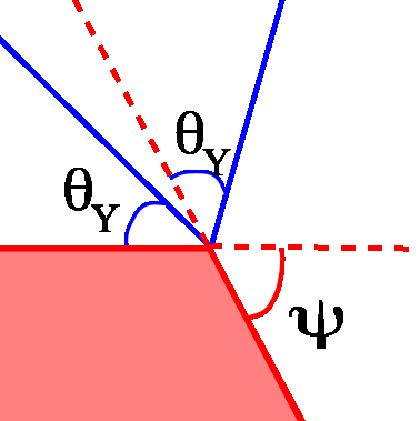}}
\subfigure[]{\label{fig:schematic}\includegraphics[width=120mm]{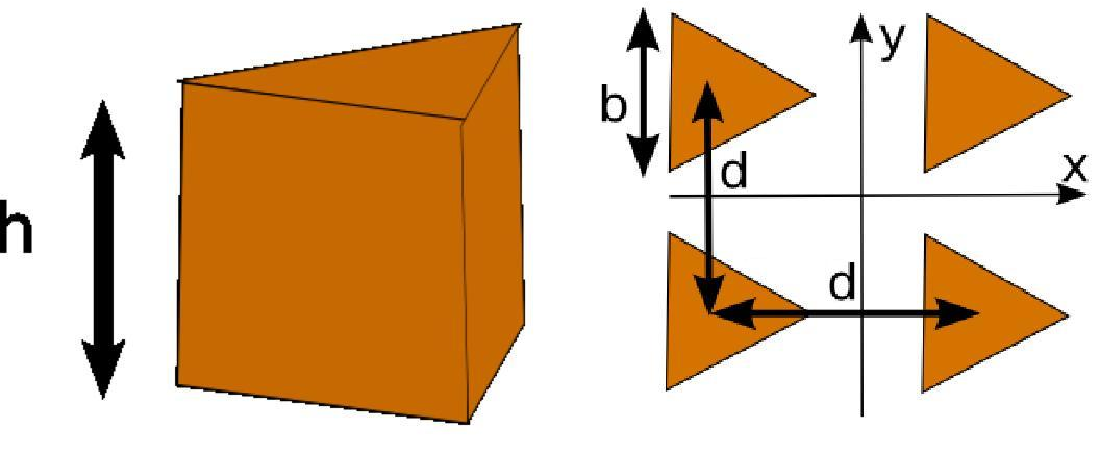}}
\caption{(a) Illustration of the Gibbs' Criterion. At a sharp corner a
wetting contact line remains pinned for all angles between $\young$ and $\young+\psi$ 
(between the full lines). (b) Diagram showing the geometry and post dimensions for the 
simulations described in Sec.~\ref{subsection:geometry}.}
\end{figure}
Contact line pinning occurs when an interface moving across a surface meets a convex 
corner. The criterion for pinning, proposed by Gibbs~\cite{Gibbs} and demonstrated 
experimentally by Oliver {\it et al.}~\cite{Oliver}, is that the contact angle can 
take a range of values spanning the dihedral angle of the corner, as shown in 
Fig.~\ref{fig:gibbs}. Over this range, the contact angle with respect to the dry plane 
is too low for the contact line to advance, and that with the wet plane is too high 
for the contact line to recede. Pinning on surface features can lead to the threshold 
angle for imbibition being substantially lower than predicition (\ref{eqn:bico}), as 
was demonstrated by Courbin et al~\cite{Courbin,Courbin2}. Furthermore it has been 
shown that anisotropic surface features lead to anisotropic spreading~\cite{Kim,Chu}. 
In these proceedings, we use the lattice Boltzmann method to study imbibition through 
an array of posts with uniform polygonal cross-section, building on previous 
work~\cite{BlowKusumaatmaja}. We show that the mechanism of contact line pinning 
differs with direction, and explain how this leads to anisotropic spreading.

\section{Simulation approach}
\label{section:LBM}
We model the sytem as a diffuse-interface, two-phase fluid in contact with a solid 
substrate. The thermodynamic state of the fluid is described by an order parameter 
$\rho(\mathbf{r})$, corresponding to the density of the fluid at each point 
$\mathbf{r}$. The equilibrium properties are modelled by a Landau free energy 
functional over the spatial domain of the fluid $\mathcal{D}$, and its boundary with 
solid surfaces $\partial\mathcal{D}$,
\begin{equation}
\Psi=\iiint_{\mathcal{D}}\left(p_{\mathrm{c}}\left\{\nu^{4}-2\beta\tau_{\mathrm{w}}(1-
\nu^{2})-1\right\}-\mu_{\mathrm{b}} \rho+\tfrac{1}{2}\kappa\vert\nabla 
\rho\vert^{2}\right)dV-\iint_{\partial \mathcal{D}}\mu_{\mathrm{s}}\rho dS\;.         
\label{eqn:fluidFreeEnergy}
\end{equation}
The first term in the integrand of (\ref{eqn:fluidFreeEnergy}) is the bulk free energy 
density, where $\nu=(\rho-\rho_{\mathrm{c}})/\rho_{\mathrm{c}}$ and 
$\rho_{\mathrm{c}}$, $p_{\mathrm{c}}$, and $\beta\tau_{\mathrm{w}}$ are constants. It 
allows two equilibrium bulk phases, liquid and gas, with 
$\nu=\pm\sqrt{\beta\tau_{\mathrm W}}$. The second term is a Lagrange multiplier 
constraining the total mass of the fluid. The third term is a free energy cost 
associated with density gradients. This allows for a finite-width, or diffuse, 
interface to arise between the bulk phases, with surface tension 
$\gamma=\tfrac{4}{3}\rho_{\mathrm{c}}\sqrt{2\kappa p_{c}(\beta\tau_{\mathrm{w}})^{3}}$ 
and width 
$\chi=\tfrac{1}{2}\rho_{\mathrm{c}}\sqrt{\kappa/(\beta\tau_{\mathrm{w}}p_{\mathrm{c}})
}$. The boundary integral takes the form proposed by Cahn~\cite{Cahn}. Minimising the 
free energy leads to a Neumann condition on the density
\begin{equation}
\partial_{\perp}\rho = -\mu_{\mathrm{s}}/\kappa\;.        \label{eqn:cahn}
\end{equation}
The wetting potential $\mu_{\mathrm{s}}$ related to the 
$\young$ of the substrate by~\cite{Briant}
\begin{equation}
\mu_{\mathrm{s}} = 2\beta\tau_{\mathrm{w}}\sqrt{2p_{\mathrm{c}}\kappa}
\mathrm{sign}\left(\tfrac{\pi}{2}-\young\right)\sqrt{\cos{\tfrac{\alpha}{3}}
\left(1-\cos{\tfrac{\alpha}{3}}\right)}\;,\;\;\;\;
\alpha=\arccos{(\sin^2{\theta_Y})}\;.\label{eqn:youngAngleToPotential}
\end{equation}
The hydrodynamics of the fluid is described by the continuity and the Navier-Stokes 
equations
\begin{align}
    \partial_{t}\rho+\partial_{\alpha}(\rho u_{\alpha})&=0\;,                              
\label{eqn:continuity}  \\
    \partial_{t}(\rho u_{\alpha})+\partial_{\beta}(\rho u_{\alpha}u_{\beta})&=- 
\partial_{\beta}P_{\alpha\beta}+ 
\partial_{\beta}\left(\rho\eta\left\{\partial_{\beta}u_{\alpha} + 
\partial_{\alpha}u_{\beta}\right\}+\rho\lambda\delta_{\alpha\beta}\partial_{\gamma}u_{
\gamma}\right)\;,  \label{eqn:navierStokes}
\end{align}
where $\mathbf{u}$ is the local velocity, $\mathbf{P}$ is the pressure tensor derived 
from the free energy functional (\ref{eqn:fluidFreeEnergy})
and $\eta$
and $\lambda$
are the shear and bulk kinematic viscosities respectively. 
A free energy lattice Boltzmann algorithm is used to numerically solve 
Eqns.~(\ref{eqn:continuity},\ref{eqn:navierStokes})~\cite{Yeomans,Succi,Swift}. At the substrate we impose the boundary condition (\ref{eqn:cahn})~\cite{Briant,Dupuis}, and a condition of no-slip~\cite{Ladd,Pooley,Bouzidi}.

We choose $\kappa=0.01$, $p_{\mathrm{c}}=0.125$, $\rho_{\mathrm{c}}=3.5$, 
$\tau_{\mathrm{W}}=0.3$ and $\beta=1.0$, giving an interfacial thickness $\chi=0.9$, 
surface tension $\gamma=0.029$ and a density ratio of $3.42$. The viscosity ratio is 
$\eta_{\mathrm{L}}/\eta_{\mathrm{G}}=7.5$.

\section{Identifying the pinning mechanisms}
\label{section:pinningMechanism}
\label{subsection:geometry}

We consider a rectangular array of posts on a flat substrate. The cross-section of 
each post is uniform, and is an equilateral triangle, oriented to point along a 
primary axis of the array taken to be the $x$-direction (see 
Fig.~\ref{fig:schematic}). In our simulations we hold the array spacing $d$ and post 
side-length $b$ at $40$ and $20$ lattice units respectively and vary the post height 
$h$. We find that rescaling the system such that $d=60$ does not change the threshold 
angles of spreading significantly{\footnote{Reducing the system size to $d=20$ leads to slightly lower values
for the depinning thresholds which explains the small quantitative differences to the
results we present in \cite{BlowKusumaatmaja}}}. The posts and substrate are taken to
have the same Young angle $\young$.

We consider the advance of a straight contact line, which is parallel to the $y$-axis. 
We exploit periodic boundary conditions and use a simulation box of length $d_{y}$ 
along $y$. We further halve the computational burden by taking $x=0$ as a plane of 
reflectional symmetry, and we compare the dynamics for triangles pointing away from, 
or towards, the origin. To simulate imbibition fed by a mother drop resting on the 
surface would require a very large simulation box, and be prohibitively costly in 
terms of computer time. Since we are only interested in the details of flow amoungst 
the posts, we instead feed imbibition from a `virtual reservior', a small region $\sim 
6$ lattice points wide spanning the centre of the box where $\nu$ is fixed to 
$\sqrt{\beta\tau_{\mathrm{w}}}$ at each time step of the simulation. In this way, 
liquid is introduced while there is outwards flow, but once the interface is fully 
pinned, no new liquid enters the system.

$\young$ is then decreased quasistatically, and we record the value at which depinning 
and spreading to the next post occurs. For the geometry we describe, 
Eqn.~(\ref{eqn:bico}), which describes imbibition with no pinning, gives an upper 
bound of the threshold angle of~\cite{Bico}
\begin{equation}
\sec\youni=1+\frac{12bh}{4d^{2}-\sqrt{3}b^{2}}=1+\frac{12h/b}{16-\sqrt{3}} 
\label{eqn:bicoTriangles}\;.
\end{equation}

\subsection{Pinning of a connected interface}
\label{subsection:CCL}
\begin{figure}
\centering
\subfigure[]{\label{fig:CCLsnapshots}\includegraphics[width=75mm]{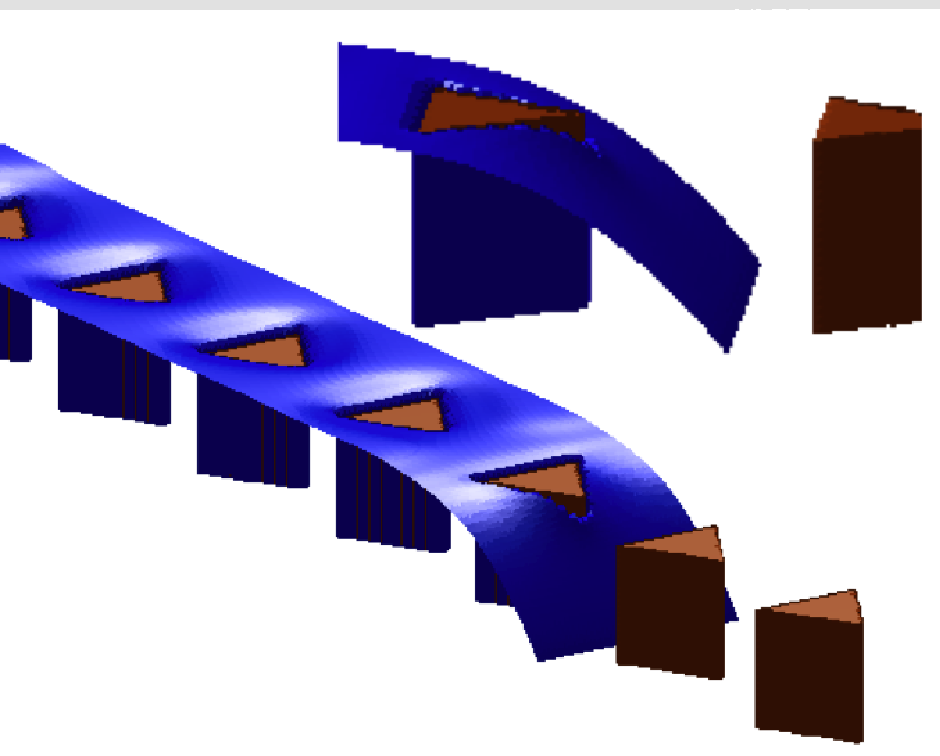}}
\subfigure[]{\label{fig:DCLsnapshots}\includegraphics[width=105mm]{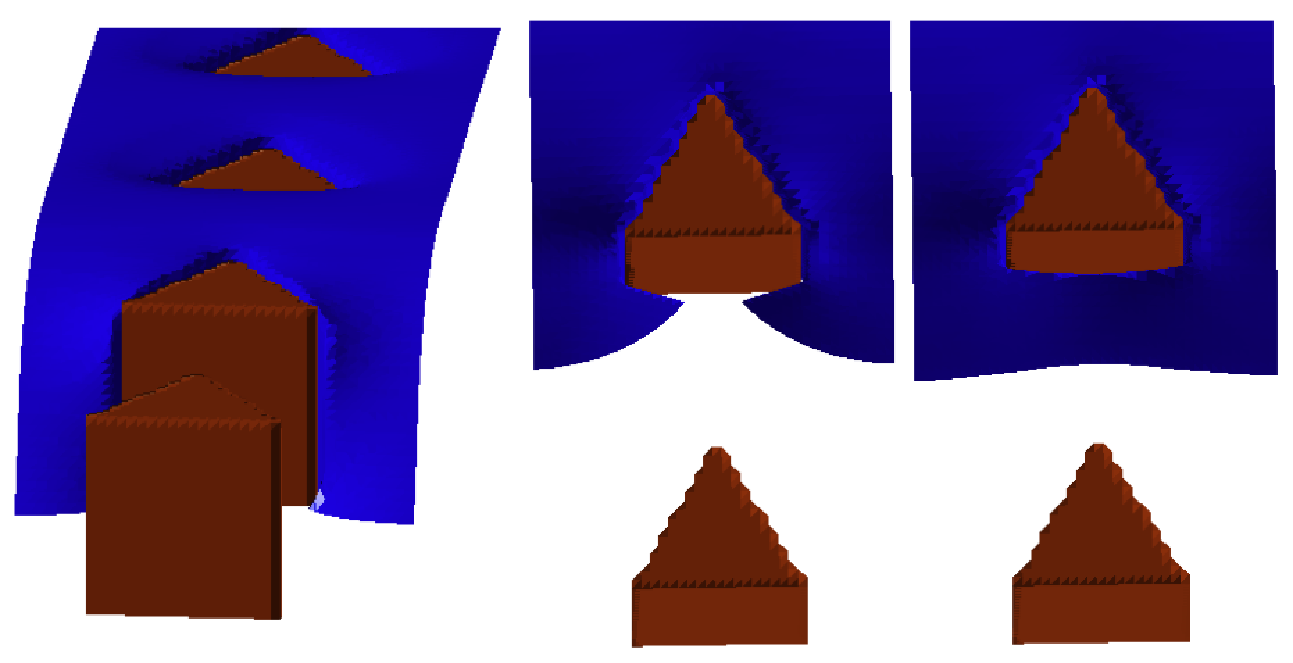}}
\caption{Snapshots of (a) the {\it connected contact line} and (b) the {\it 
disconnected contact line} mechanisms of pinning. In the middle image of (b) a gap 
appears between the interface and the face of the post. This is because the liquid 
(blue) surface represents the density $\rho_{\mathrm{c}}$, and close to the concave 
corner the wetting potential
increases the density above $\rho_{\mathrm{c}}$ }
\label{fig:snapshots}
\end{figure}
\begin{figure}
\centering
\includegraphics[width=150mm]{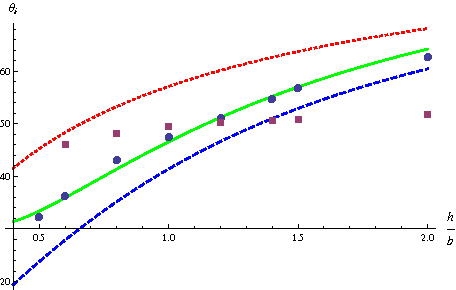}
\caption{The threshold angle for depinning along $+x$ (indigo circles) and $-x$ (mauve 
squares) as a function of $h/b$. Depinning in these directions occurs according to the 
connected and disconnected contact line mechanisms respectively. The predictions from 
Eqns.~(\ref{eqn:bicoTriangles}) in red (dotted), (\ref{eqn:courbinTriangles}) in blue 
(dashed), and (\ref{eqn:courbinMod}) in green (full) are plotted for comparison.}
\label{fig:resultsGraph}
\end{figure}
Snapshots showing one pinning mechanism, for a film advancing in the direction of the 
points of the triangles, are shown in Fig.~\ref{fig:CCLsnapshots}. The film is of 
height $h$ up to the leading triangle, and then descends with increasing $x$ to meet 
the substrate at the Young angle. There are two ways in which the contact line can 
move forward. Firstly, it could make a shallower angle at the substrate, but this 
would increase the free energy away from the minimum characterised by 
Eqn.~(\ref{eqn:young}). Secondly, the top of the film could move forwards, but this 
would create liquid-gas interface and hence also have a free energy cost.

Depinning will occur when $\young$ is sufficiently small that the contact line on the 
base reaches the next post. This depinning pathway, which we shall term the {\it 
connected contact line} mechanism, was elucidated by Courbin et 
al\cite{Courbin,Courbin2}, who showed that
\begin{equation}
\tan\youni=\frac{\text{post height}}{\text{post spacing}}  = 
\frac{h}{d-\tfrac{\sqrt{3}}{2}b}=\frac{h/b}{2-\tfrac{\sqrt{3}}{2}}\;.\label{eqn:courbi
nTriangles}
\end{equation}
Numerical results for the variation of the depinning angle with $h/b$ are shown in 
Fig.~\ref{fig:resultsGraph} as indigo circles, and 
Eqns.~(\ref{eqn:bicoTriangles},~\ref{eqn:courbinTriangles}) are plotted as red and 
blue curves respectively. Comparing the simulation data to the blue curve, we see that 
the simulation values are significantly higher than those predicted, expecially for 
lower values of $h/b$.

To resolve the discrepency we note that Eqn.~(\ref{eqn:courbinTriangles}) assumes a flat interface. A positive Laplace pressure $\Delta p$ 
will instead produce a convex curvature, enabling the interface to extend further 
across the substrate. Neglecting curvature in the $y$ 
direction, we model the interface in the $xz$ plane as a circular arc with radius of 
curvature $R=\gamma/\Delta p$ given by Laplace's law. The contact angle with the 
substrate will then be modified to
\begin{equation}
\tan\left(\youni-\beta\right)=\tfrac{h}{s}\;,
\end{equation}
where $\beta$ is the angle of bulge, given by $\sqrt{h^{2}+s^{2}}=2R\sin\beta$. The depinning threshold is thus given by
\begin{equation}
\youni=\arctan\left[\tfrac{h}{s}\right]+
\arcsin\left[\tfrac{\sqrt{h^{2}+s^{2}}}{2R}\right]\;.       \label{eqn:courbinMod}
\end{equation}
We expect the dominant contribution to the Laplace pressure to result from confinement 
in the $z$ direction. 
Therefore, we shall assume $\Delta p \propto h^{-1}$. Writing $R=Ah$ and $s=d-Bb$, a 
least squares fit of the data to Eqn.~(\ref{eqn:courbinMod}), with respect to $A$ and 
$B$, was performed. The optimisation found $B=0.822$, barely different from the value 
$\tfrac{\sqrt{3}}{2}\approx0.866$ used in Eqn.~(\ref{eqn:bicoTriangles}), and 
$A=7.09$. Eqn.~(\ref{eqn:courbinMod}) is plotted with these coefficients as the green 
curve in Fig.~\ref{fig:resultsGraph}, and the fit is very reasonable.

\subsection{Pinning of a disconnected interface}
\label{subsection:DCL}
We now present simulation results with the posts pointing towards the origin, and 
identify a second mechanism for (de)pinning, shown in Fig.~\ref{fig:DCLsnapshots}. Now 
the advancing front is disconnected, and is pinned at the vertical edges of the posts. 
The base of the film is pulled forward by the hydrophilic substrate, but there is a 
free energy cost associated with the growth of the interface as it spreads out from 
the gap. As the Young angle is quasistatically decreased, the contact line creeps onto 
the blunt faces of the posts, near to the base substrate, but remains pinned to the 
post edges at higher $z$, where the angle made between the interface and the blunt 
faces remains less than $\young$. When $\young$ becomes sufficiently small, the 
depinned parts of the contact lines from neighbouring gaps meet each other midway. 
Once connected, the interface readily wets up the posts and out across the substrate.

We shall refer to this as the {\it disconnected contact line} pinning mechanism. The 
threshold for depinning is plotted in Fig.~\ref{fig:resultsGraph} as mauve squares. 
The dependence on $h/b$ is different to that for motion along $+x$. When $h/b$ is low, 
the depinning angle closely follows the upper bound given by 
Eqn.~(\ref{eqn:bicoTriangles}), indicating that pinning by the posts is weak in this 
regime. For larger values of the ratio $h/b$, $\youni$ levels off to $\sim 
51^{\circ}$.

\section{Imbibition through polygonal posts}
\label{section:3Dimbibition}

We now present simulation results for films spreading through arrays with various 
lattice symmetries and post geometries. We use arrays which are several posts wide in 
both the $x$ and $y$ directions, such that the film is not connected over periodic 
boundaries. We again use a virtual reservoir, this time located at a small location at 
the centre of the array, but we hold $\young$ constant over time. We discern how both 
the arrangement, and the geometries, of the posts affect the dynamics of the 
interfaces, and the final film shapes. These can be interpreted in terms of the 
pinning mechanisms identified in Sec.~\ref{section:pinningMechanism}.

For our simulations we use $d=40$, $b=20$, $h=30$ and $\young=55^{\circ}$. According 
to Fig.~\ref{fig:resultsGraph}, these parameters should allow and inhibit spreading in 
the $+x$ and $-x$ directions respectively. In 
Figs.~\ref{fig:3DTriSqu},~\ref{fig:3DHexSqu} and \ref{fig:3DTriHex}, we show plan 
views of the substrate at various times in the evolution of the film. The posts are 
shown in brown, the wetted substrate in blue, and the unwetted substrate in white.
\subsection{A square array of triangular posts}
\begin{figure}
\centering
\includegraphics[width=150mm]{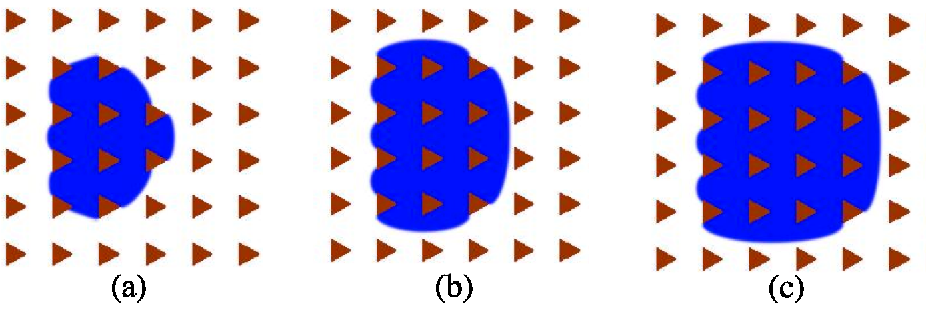}
\caption{Spreading on a square array of triangles}
\label{fig:3DTriSqu}
\end{figure}
We first consider the system studied in Sec.~\ref{section:pinningMechanism}, extended 
in the $y$ direction. The shape of the film at intermitant times is shown in 
Fig.~\ref{fig:3DTriSqu}. Advance of the film is possible in the $+x$ and $\pm y$ 
directions, via the connected contact line mechanism, but the film is barred from 
advancing in the $-x$ direction, where the disconnected contact line mechanism, which 
has a lower threshold angle, is relevant. Thus the surface acts as a microfludic 
diode. Such unidirectional behaviour is made possible by the triangular shape of the 
posts.
\subsection{A square array of hexagonal posts}
\begin{figure}
\centering
\includegraphics[width=160mm]{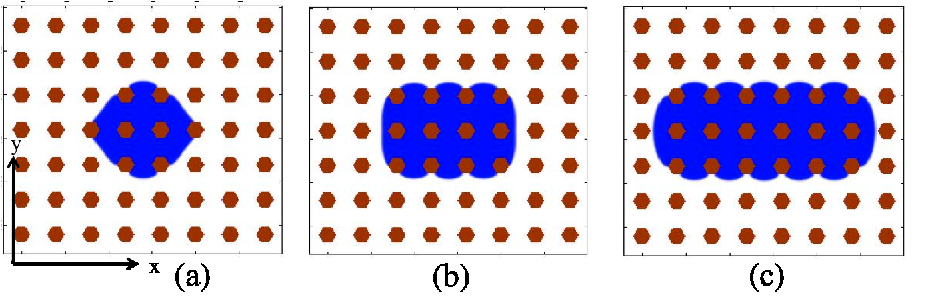}
\caption{Hexagonal posts show the connected contact line mechanism along $\pm x$ and 
the disconnected contact line mechanism along $\pm y$.}
\label{fig:3DHexSqu}
\end{figure}
Having considered exclusively triangular posts thus far in the chapter, we now turn 
our attention to posts whose cross-sections are regular hexagons. We find that the two 
depinning mechanisms discerned for triangles, in Sec.~\ref{section:pinningMechanism}, 
may also be applied to hexagons, but that their directional distribution of occurrence 
is different. We consider hexagonal posts in a square array, oriented so that the 
corners point along $\pm x$. Along these two directions, as might be expected, the 
(de)pinning behaviour follows the connected contact line mechanism. Conversely, the 
faces point along $\pm y$, and it is the disconnected contact line mechanism which 
determines the (de)pinning in these directions. Fig.~\ref{fig:3DHexSqu} shows the 
spreading of a film on a square array of hexagons. Since advance of the liquid is 
permitted along $\pm x$ but barred along $\pm y$, a stripe of fluid is formed.

\subsection{A hexagonal array of triangular posts}
\begin{figure}
\centering
\includegraphics[width=150mm]{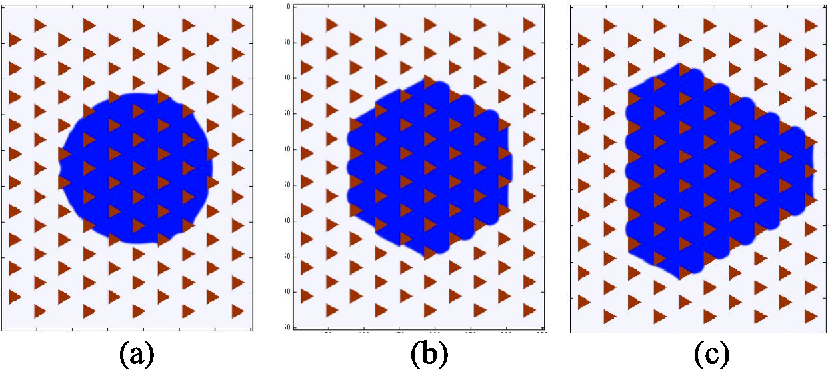}
\caption{Spreading on a hexagonal array of traingles}
\label{fig:3DTriHex}
\end{figure}

We now simulate a hexagonal lattice of posts with spacing $d=40$, and the triangles 
aligned with lattice directions, as shown in Fig.~\ref{fig:3DTriHex}. We start with a 
circular film with diameter spanning several posts (Fig.~\ref{fig:3DTriHex}(a)). As 
spreading begins, the film quickly facets into a hexagon, by aligning its sides with 
posts in the immediate vicinity (Fig.~\ref{fig:3DTriHex}(b)). Spreading continues, via 
the connected contact line mechanism, along the directions of the three corners of the 
posts, but the interface is pinned, by the disconnected contact line mechanism, along 
the faces. As a result, the facets along the corner directions shrink as they advance 
(Fig.~\ref{fig:3DTriHex}(c)).

\section{Discussion}

We have performed Lattice Boltzmann simulations of imbibition on hydrophilic 
substrates patterned with posts, whose cross-sections are regular polygons. Our 
motivation was to identify pinning mechanisms on the posts and show how these lead to 
anistropic spreading behaviour on the surface.

We began by considering the advance of a long planar front along a row of triangular 
posts. This enabled us to take advantage of periodic boundaries in the simulations, 
reducing computational expense, and to isolate particular pinning behaviours. The 
simulations showed that the critical value of $\young$ at which the interface advances 
differs between directions relative to the triangles. Hence there is a range of 
$\young$ in which spreading is unidirectional, with the exact range and direction of 
the anisotropy depending on the relative dimensions of the substrate. The cause is differing depinning routes: one where the contact line along the 
base substrate is connected, and one where it is disconnected, punctuated by the blunt 
edges of the posts.

We showed that a square lattice of triangular posts inhibits spreading in one 
direction, while if hexagonal posts are used, the spreading is bidrectional, with 
films elongating. Finally we investigated spreading amongst a hexagonal lattice of 
triangular posts. The three-fold rotational symmetry of this geometry leads to the 
formation of a triangular film.

In future work it would be of interest to consider how the spreading is affected if the post cross section changes with height, and how electrowetting might be used to locally control the contact angle, and hence the spreading characteristics~\cite{Heikenfield}. 

\begin{acknowledgements}
We thank H. Kusumaatmaja, B. M. Mognetti and R. Vrancken for helpful discussions.
\end{acknowledgements}


%

\label{lastpage}
\end{document}